\documentclass[aps,prx,twocolumn,superscriptaddress,showpacs,epsf]{revtex4-2}

\usepackage{nccmath}
\usepackage{layouts}
\usepackage{comment}
\usepackage{ulem}
\usepackage{color}
\usepackage{graphicx}% Include figure files
\usepackage{subfigure}
\usepackage{dcolumn}% Align table columns on decimal point
\usepackage{bm}% bold math
\usepackage{amsmath}
\usepackage{amsfonts}

\usepackage{braket}
\usepackage{diagbox, eqparbox, hhline}
\usepackage{hyperref}% add hypertext capabilities
\DeclareMathAlphabet{\mathpzc}{OT1}{pzc}{m}{it}
\usepackage{xcolor}
\usepackage{color,soul}
\usepackage{physics}
\newcommand{\lsim}{{\;\raise0.3ex\hbox{$<$\kern-0.75em\raise-1.1ex\hbox{$\sim$}}\;}}
\begin{document}
	
	\title{Metastable states and hidden phase slips in nanobridge SQUIDs}
	%\title{Phase dynamics of a nanoscale SQUID}
	
	\author{Lukas Nulens}
	\affiliation{Quantum Solid-State Physics, Department of Physics and Astronomy, KU Leuven, Celestijnenlaan 200D, B-3001 Leuven, Belgium}
	\author{Heleen Dausy}
	\affiliation{Quantum Solid-State Physics, Department of Physics and Astronomy, KU Leuven, Celestijnenlaan 200D, B-3001 Leuven, Belgium}
	\author{Micha{\l} J. Wyszy\'nski}
	\affiliation{Theory of Functional Materials, Department of Physics, University of Antwerp, Groenenborgerlaan 171, B-2020 Antwerp, Belgium}
	\author{Bart Raes}
	\affiliation{Quantum Solid-State Physics, Department of Physics and Astronomy, KU Leuven, Celestijnenlaan 200D, B-3001 Leuven, Belgium}
	\author{Margriet J. Van Bael}
	\affiliation{Quantum Solid-State Physics, Department of Physics and Astronomy, KU Leuven, Celestijnenlaan 200D, B-3001 Leuven, Belgium}
	\author{Milorad V. Milo\v{s}evi\'c}
	\affiliation{Theory of Functional Materials, Department of Physics, University of Antwerp, Groenenborgerlaan 171, B-2020 Antwerp, Belgium}
	\author{Joris Van de Vondel}
	\affiliation{Quantum Solid-State Physics, Department of Physics and Astronomy, KU Leuven, Celestijnenlaan 200D, B-3001 Leuven, Belgium}
	\email{joris.vandevondel@kuleuven.be}

	\date{\today}
	
	\begin{abstract}
		We fabricated an asymmetric nanoscale SQUID consisting of one nanobridge weak link and one Dayem bridge weak link. The current phase relation of these particular weak links is characterized by multivaluedness and linearity. While the latter is responsible for a particular magnetic field dependence of the critical current (so-called vorticity diamonds), the former enables the possibility of different vorticity states (phase winding numbers) existing at one magnetic field value. In experiments the observed critical current value is stochastic in nature, does not necessarily coincide with the current associated with the lowest energy state and critically depends on the measurement conditions. In this work, we unravel the origin of the observed metastability as a result of the phase dynamics happening during the freezing process and while sweeping the current. Moreover, we employ special measurement protocols to prepare the desired vorticity state and identify the (hidden) phase slip dynamics ruling the detected state of these nanodevices. In order to gain insights into the dynamics of the condensate and, more specifically the hidden phase slips, we performed time-dependent Ginzburg-Landau simulations.

	\end{abstract}
	
	%\pacs{Valid PACS appear here}% PACS, the Physics and Astronomy Classification Scheme.
	%\keywords{Valid keywords appear here}%Use showkeys class option if keyword display desired
	
	\maketitle
	
	\section{Introduction} \label{sec.intro}
	% Wat zijn phase slips en waarom zijn ze interesant/relevant?
	Phase slips - topological fluctuations of the order parameter - are an indispensable ingredient for understanding the behavior of various superconducting nanodevices \cite{article_mooij2005phase,article_baumans2017statistics}. In a one-dimensional superconducting nanowire, quantum and/or thermal phase slip events are responsible for the onset of a dissipative state \cite{article_lau2001quantum,article_li2011switching,article_aref2012quantitative}. Nevertheless, in order to detect these events using a dc measurement a sufficient high phase slip rate has to be induced and/or a local hotspot has to be created \cite{article_berdiyorov2014dynamics}. A more pronounced impact of a single phase slip event can be expected in ring like structures \cite{article_Petkovic_I_2016,article_kenawy2020electronically}. This stems from the existence of different metastable states at one magnetic field value, where each state corresponds to a unique value of the winding number of the superconducting order parameter along the ring - the vorticity, $n_v$. Even a single phase slip event modifies the vorticity of the ring, which is directly linked to a variety of macroscopic observables \cite{article_sahu2009individual,article_singh2013observation}. The ability to detect the impact of a single phase slip event was recently used to fabricate a persistent Josephson phase-slip memory cell with topological protection \cite{article_giazotto_2021}. 
	\newline
	\newline
	% Dan overgaan naar nanosquids als interessantie applicatie van /studieobject voor phase slips. Laten weten hoe ze impact kunnen 'voelen' en hoe deze geobserveerd kan worden. 
	In case of a Superconducting Quantum Interference Device (SQUID), which is in essence a superconducting loop with two weak links, the aforementioned ingredients are also present. Each vorticity state is characterized by a unique critical current versus field, $I_c(B)$, dependence. Therefore it offers a simple readout method to identify the vorticity state of the device. Metastability can be induced through the use of a long nanobridge as one of the weak links \cite{article_hazra2019nanobridge}. By controlling the length and width of this nanobridge, the kinetic inductance and hence the shape of the $I_c(B)$-oscillations of the device can be controlled \cite{article_dausy2021impact}. Moreover, the energy barrier between different vorticity states can be tuned by changing the applied magnetic flux or the applied bias current. When the energy barrier approaches zero, a stochastic or deterministic phase slip process induces a transition to either a dissipative state or to another vorticity state. The ability to read and write the flux state of a nanoSQUID under well chosen biasing conditions can be used to design a flux based memory \cite{article_Murphy_A_2017,article_Ilin_2021}.
	\newline
	\newline
	% Focus hoe in de nanosquid dit inderdaad gemeten wordt.
	In order to obtain the critical current of a device (and as such the vorticity state), the voltage is probed while performing a sweep from a large positive (or negative) bias current towards a large negative (or positive) bias current, for a given external magnetic field. These experiments have demonstrated that the obtained critical current value (and thus also the final vorticity state) fluctuates around multiple discrete values \cite{article_murphy2017asymmetric,article_collienne2021nb}. This behavior clearly originates from the metastability of different vorticity states and the stochastic nature of the phase slip events. If one wants to use these devices for memory based applications it is of utmost importance to understand how the vorticity state preparation and readout are governed by the underlying phase dynamics.   
	\newline
	\newline
	% Wat hebben wij gedaan. 
	In this work we use the unique properties of a nanobridge SQUID (NBSQUID) to initialize it in a particular vorticity state \cite{article_murphy2017asymmetric}. This NBSQUID provides an interesting platform to investigate the metastability of the vorticity states and the impact of phase slips on the manifestation of particular critical current branches. We observe that different branches manifest in the $I_c(B)$ characteristic measured by an $IV$-sweep following the initialization of the NBSQUID in the $n_v=0$ state at zero field, than when measured after an uncontrolled state preparation. In particular, we observe that for a limited field range around zero field, only the $n_v=0$ branch is probed: the $I_c(B)$ is single valued. For larger field values the $I_c(B)$ becomes multivalued again. This indicates that the NBSQUID has altered its vorticity state due to a stochastic phase slip event when crossing the vorticity diamond of the $n_v=0$ state, which indicates the presence of so-called hidden phase slips. To the best of our knowledge, the impact of the initial state on the multivaluedness of the $I_c(B)$ characteristic of a NBSQUID has not been demonstrated before. As the vorticity state is known only after the transition to a dissipative state, the previous methods do not allow to investigate the vorticity of the NBSQUID below the bias current necessary to induce this transition. In order to gain information about the stability of the $n_v=0$ state at low bias currents, we employed a measurement protocol \cite{article_Murphy_A_2017} which relies on the unique properties of an asymmetric NBSQUID to prepare and read out the vorticity state. We demonstrate that the NBSQUID remains trapped in the $n_v=0$ state within the whole so called `vorticity diamond' region, even though other vorticity states have lower energies in this region. Finally, time-dependent Ginsburg-Landau (tdGL) simulations of very similar device geometries gave a close agreement to the observed dynamics and allowed us to explore the hidden phase dynamics inaccessible to the experiment. 
	%----------------------------------------
	
	\begin{figure*}[ht!]
		\centering
		\subfigure{\includegraphics[width=\linewidth]{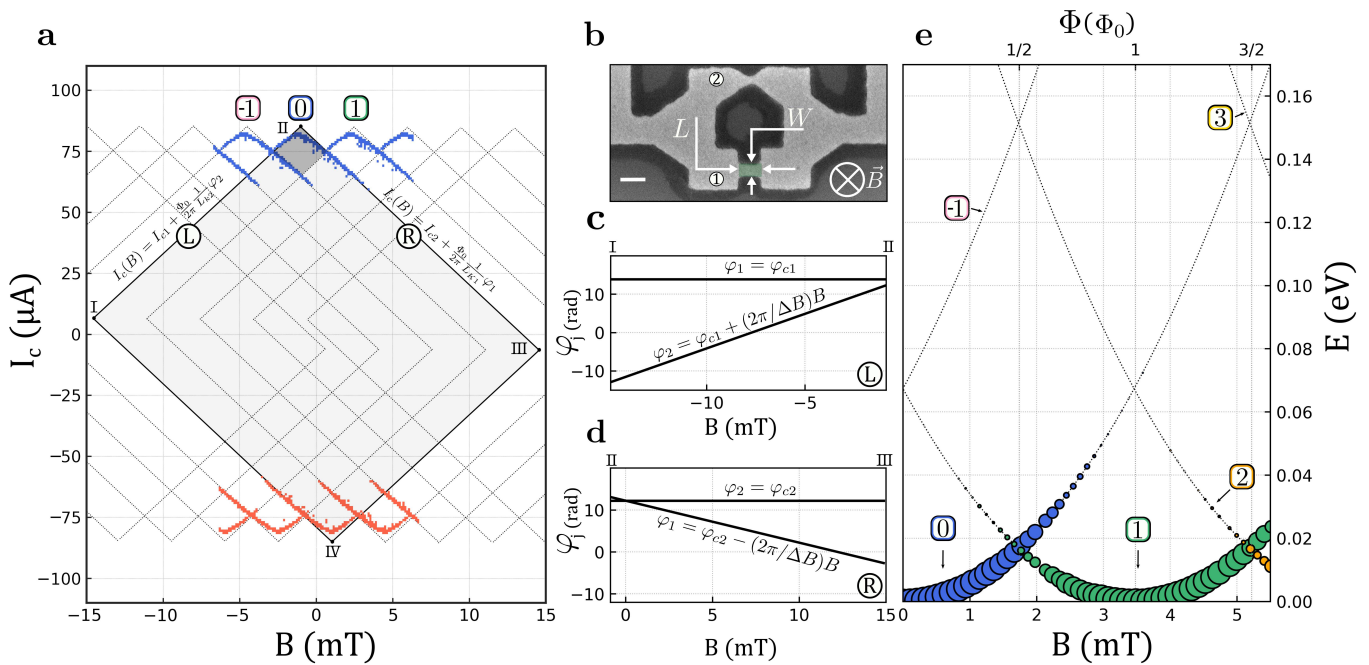}}
		\caption{(a) The critical currents as a function of the applied magnetic field. The measured critical currents for positive (negative) bias current are shown in blue (red). The dashed lines represent the vorticity diamonds generated by
			the model discussed in the text. The vorticity number is indicated by a number in a rounded square. The $n_v=0$ diamond is indicated by a light gray fill and a solid edge. The $n_v=0$ unique vorticity diamond (UVD) has a dark gray fill and correpsonds with the phase space where only the $n_v=0$ state exists. (b) False colored Scanning Electron Microscopy (SEM) image of a prototypical NBSQUID device as studied here. The top junction is a Dayem bridge, while the bottom junction indicated in green is a nanobridge with length $L$ and width $W$. The white scale bar represents 200 nm. When measuring flux oscillations, the applied field $B$ is oriented as shown. (c),(d) The evolution of the phases $\varphi_j$ with $j = {1,2}$ along the branches indicated by $L$ and $R$ respectively. Here $j=1$ denotes the nanobridge SQUID arm, and $j=2$ the Dayem bridge SQUID arm. (e) The energy of the different vorticity states of the SQUID under investigation as a function of the applied field (or the equivalent applied flux) at zero bias current. The size of the colored dots represents the probability to measure the critical current corresponding to the indicated vorticity state for positive bias currents, extracted from the experimental data in panel a.}
		\label{fig1}
	\end{figure*}

	\section{Multiple metastable vorticity states in a NBSQUID}\label{sec.model}
	
	Figure \ref{fig1}b shows a scanning electron microscopy (SEM) image of a similar NBSQUID device as studied in this work. The SQUID contains one Dayem bridge (top junction) and one nanobridge weak link (bottom junction). For the device studied in this work, the nanobridge dimensions are determined from SEM imaging as length $L = 176$ nm and width $W = 54$ nm. The sample is fabricated using conventional electron beam lithography, followed by pulsed laser deposition of a 25 nm/5 nm thick MoGe/Au film and a standard lift-off process. The device has a superconductor-to-normal-state transition temperature of approximately $T_c \approx 6$ K. From measurements of the superconducting-to-normal-state phase boundary on similarly prepared plain films of MoGe/Au, the coherence length can be determined as approximately $\xi(T=0$ K) $\approx$ 10 nm \cite{article_Motta_M_2014}.
	\newline
	\newline
	Figure \ref{fig1}a shows the measured critical current versus field data $I_c (B)$ of the NBSQUID obtained at $T$ = 300 mK. The $I_c (B)$ is obtained in the conventional measurement method, meaning that at each magnetic field value, a set of 150 $IV$ measurements in both current sweep directions is acquired using a current ramp rate of 1.8 mA/s. Note that the $IV$-curves of the NBSQUIDS are highly hysteretic as the transition from the superconducting state to the normal state is dominated by Joule heating. For each of these curves the critical current is extracted and shown in Figure \ref{fig1}a as a small dot. The blue (red) dots indicate the field dependence of the critical current obtained when sweeping the current from a large negative (positive) bias current of $\mp$ 90 $\mu$A towards a large positive (negative) bias current of  $\pm$ 90 $\mu$A. The observed oscillation period is $\Delta B=3.48$ mT, which agrees with the value expected from geometrical considerations. Thermal or quantum fluctuations can cause a premature escape from the superconducting state before the depairing current is reached, resulting in a stochastic distribution of the critical current around an average value \cite{article_baumans2017statistics}. The solid black lines result from a fit of the $I_c (B)$ data to the model introduced in Ref. \cite{article_Murphy_A_2017}.
	\newline
	\newline
	This model considers a SQUID containing two weak links, which both have a linear current-phase relationship (C$\Phi$R):
	%------------------------------------------------------------
	\begin{equation}\label{eq.CPR_kinetic}
		I_j = I_{cj} \frac{\varphi_j}{\varphi_{cj}}=\frac{\Phi_0}{2\pi}\frac{ 1}{L_{Kj}} \varphi_j .
	\end{equation}
	%------------------------------------------------------------
	Here $I_j$ represents the supercurrent through the $j$-th weak link, with $j = {1,2}$ and $\varphi_j$ is the phase difference of the macroscopic wavefunction taken between the end points of the $j$-th weak link. Further, $I_{cj}\geq0$ is the critical current and $\varphi_{cj}\geq0$ is the critical phase difference at which the weak link switches to the dissipative state. It is customary to introduce the kinetic inductance $L_{Kj}=\left(\varphi_{cj}/I_{cj}\right) \left(\Phi_0/2\pi\right)$ of the $j$-th arm. Note that an almost linear C$\Phi$R has been predicted for thin and long wires ($L>3\xi$) \cite{article_Hasselbach_K_2002}. The current through the $j$-th arm can be determined from the condition that the total current through the SQUID is given by: 
	
	%------------------------------------------------------------
	\begin{equation}\label{eq.kirchoff}
		I_{bias}=\sum_{j=1,2} I_j =\sum_{j=1,2}\frac{\Phi_0}{2\pi}\frac{ 1}{L_{Kj}} \varphi_j ,
	\end{equation}
	%------------------------------------------------------------
	and the fact that the order parameter must be single valued. This means that the gauge invariant phase differences around the loop must add up to an integer multiple of $2\pi$:
	%------------------------------------------------------------
	\begin{equation}\label{eq.parks}
		\varphi_1-\varphi_2+2\pi \frac{B}{\Delta B}=2\pi n_v .
	\end{equation}
	%------------------------------------------------------------
	Here $\Delta B$ is the Little-Parks oscillation period and the phase difference over each wire is limited to the critical phase difference $\varphi_{cj}$. For the device studied in this work, $j=1$ denotes the nanobridge SQUID arm while $j=2$ denotes the Dayem bridge arm. As we assume that the contribution from the geometric inductance of the SQUID ($\sim 2pH\ll L_{Kj}$) to $B$ can be neglected, Equation \eqref{eq.kirchoff} and Equation \eqref{eq.parks} effectively decouple. Combining Equations \eqref{eq.kirchoff}-\eqref{eq.parks}, and the requirement that superconductivity should be destroyed if $|\varphi_j|>\varphi_{cj}$ in any of the bridges, one can calculate the total critical current of the NBSQUID for a given vorticity $n_v$ and applied magnetic field $B$. The total critical current of the SQUID $I_c (B,n_v)$ equals the smallest total applied current at which the current across either wire reaches its critical value. For each value of $n_v$ the solution for $I_c (B,n_v)$ forms a so-called vorticity diamond as shown for $n_v=0$ by a light gray fill in Figure \ref{fig1}a. For the branches $L$ and $R$ indicated in Figure \ref{fig1}a, the magnetic field dependence of the phase differences $\varphi_j$ over the two weak links are shown in panels c and d. Here $\varphi_1$ denotes the phase difference over the nanobridge, while $\varphi_2$ denotes that over the Dayem bridge.
	\newline
	\newline
	%------------------------------------------------------------
	%\begin{widetext}
	%------------------------------------------------------------
	%\begin{subequations}\label{eq.branches}
	%\begin{align}
	%A:\hspace{3mm}& \varphi_1=\varphi_{c1},\hspace{3mm} \varphi_2 (B)=\varphi_{c1}+2\pi \frac{B}{\Delta B} ,\hspace{3mm} %I_c (B)=I_{c1}+(\frac{2\pi}{\Phi_0}  L_{K2} )^{-1} \varphi_2
	%,\label{equationa}
	%\\
	%B:\hspace{3mm}& \varphi_2=\varphi_{c2},\hspace{3mm} \varphi_1 (B)=\varphi_{c2}-2\pi \frac{B}{\Delta B} ,\hspace{3mm} %I_c (B)=(\frac{2\pi}{\Phi_0}  L_{K1} )^{-1}\varphi_1 +I_{c2} ,\label{equationb}
	%\end{align}
	%\end{subequations}
	%------------------------------------------------------------
	%\end{widetext}
	%------------------------------------------------------------
	%As the $n_v=0$ vorticity diamond is point-symmetric around the origin, similar expressions exist for the corresponding %opposite branches.
	For the left ($L$) and right ($R$) branch indicated in Figure \ref{fig1}a the NBSQUID’s transition to the normal state corresponds to weak link $j=1(2)$ reaching its critical current (corresponding with a critical phase difference $\varphi_1=\varphi_{c1}$ ($\varphi_2=\varphi_{c2}$)). The $n_v$-th vorticity diamond is identical to the $n_v=0$ diamond, but shifted along the magnetic field axis by $B=n_v \Delta B$. As the vorticity diamond extends over a range $B=\varphi_c  \Delta B/\pi$, where $\varphi_c=\varphi_{c1}+\varphi_{c2}$, diamonds of adjacent vorticities overlap for $\varphi_c>\pi$ (i.e. twice the critical phase difference of $\pi/2$ of a conventional tunnel junction), resulting in a multivalued critical current.
	\newline
	\newline
	It is clear that the model captures the $I_c (B)$ characteristics well. From the fit, the physical parameters i.e. the kinetic inductance, $L_{K1, K2}$, the critical phase difference, $\varphi_{c1, c2}$, and the critical current, $I_{c1,c2}$ can be obtained. The obtained values for the device studied here are $L_{K1}=98$ pH, $L_{K2}=100$ pH, $\varphi_{c1} = 13.9$ rad, $\varphi_{c2} = 12.2$ rad, $I_{c1} = 46.4$ $\mu$A and $I_{c2} = 40.1$ $\mu$A. As can be seen from the fit, multiple vorticity states exists at each magnetic field value as $\varphi_{c1}+\varphi_{c2}>\pi$. Despite their existence, the experiment only probes the critical current of vorticity states that have a critical current value that exceeds a value of approximately $I>64$ $\mu A$. For $I>64$ $\mu A$ the sample transits immediately to the normal state when reaching the critical current of a particular vorticity state. As such, we can observe up to three critical current branches at one magnetic field value. The multivaluedness and the distribution of the critical current probed at a particular field value reflect the stochastic nature of the thermal and quantum fluctuations during the $IV$-measurement: as the current is swept from $\mp 90$ $\mu A$ to $\pm 90$ $\mu A$, the SQUID first transits from the normal to the superconducting state at the re-trapping current of $\mp 21.5$ $\mu A$. At this transition, the aforementioned thermal and quantum fluctuations are important and impact the vorticity state which the NBSQUID gets frozen into. The stochasticity of this freezing process and the resulting vorticity state initialization are reflected in the multivaluedness and the spread of the critical current probed at a particular field value.
	\newline
	\newline
	The energy stored in the NBSQUID at a given magnetic field value and current bias is given by:
	%------------------------------------------------------------
	\begin{widetext}
		%------------------------------------------------------------
		%------------------------------------------------------------
		\begin{equation}\label{eq.energy}
			E=\frac{1}{2}\sum_{j=1,2}L_{Kj}I_j^2 = \frac{1}{2}\frac{L_{K1}L_{K2}}{L_{K1}+L_{K2}}I_{bias}^2+\frac{1}{2}\frac{1}{L_{K1}+L_{K2}}\left(\frac{B}{\Delta B}\Phi_0-\Phi_0 n_v\right)^2     ,
		\end{equation}
		%------------------------------------------------------------
		%------------------------------------------------------------
	\end{widetext}
	%------------------------------------------------------------
	and is quadratic in both the bias current and the applied magnetic field. Figure \ref{fig1}e shows the energy stored in the NBSQUID, $E(n_v,B,I)$ calculated according to Equation \eqref{eq.energy}, for vorticity states $n_v=-1$, $0$, $1$, $2$ and $3$ at zero bias current as a function of the external magnetic field (or equivalently, flux). Note that an applied bias current just shifts the energy levels along the energy axis by a vorticity independent factor. At 2 mT, the energy difference between the $n_v=0$ and $n_v=1$ state is about $\sim 0.01$ eV. The size of the colored dots represents the probability to measure the critical current corresponding to the indicated vorticity state, extracted from the experimental data in panel a. It is clear that the $n_v=1$ state corresponds to the lower energy state in the range $\Phi_0/2<\Phi<3\Phi_0/2 $. Nevertheless, Figure \ref{fig1}a and e show that critical current and energy values corresponding to the $n_v=0$ and $n_v=2$ states are also observed. This indicates that the NBSQUID can be rendered in a metastable state during the freezing process, which is not necessarily the lowest in energy. The NBSQUID can remain in this metastable state because state relaxation requires the density of Cooper pairs to be locally suppressed so that the superconducting order parameter phase can exhibit a $2\pi$ discontinuity - a phase slip \cite{article_langer1967intrinsic,article_mccumber1970time}.

	\begin{figure*}[ht!]
		\centering
		\subfigure{\includegraphics[width=\linewidth]{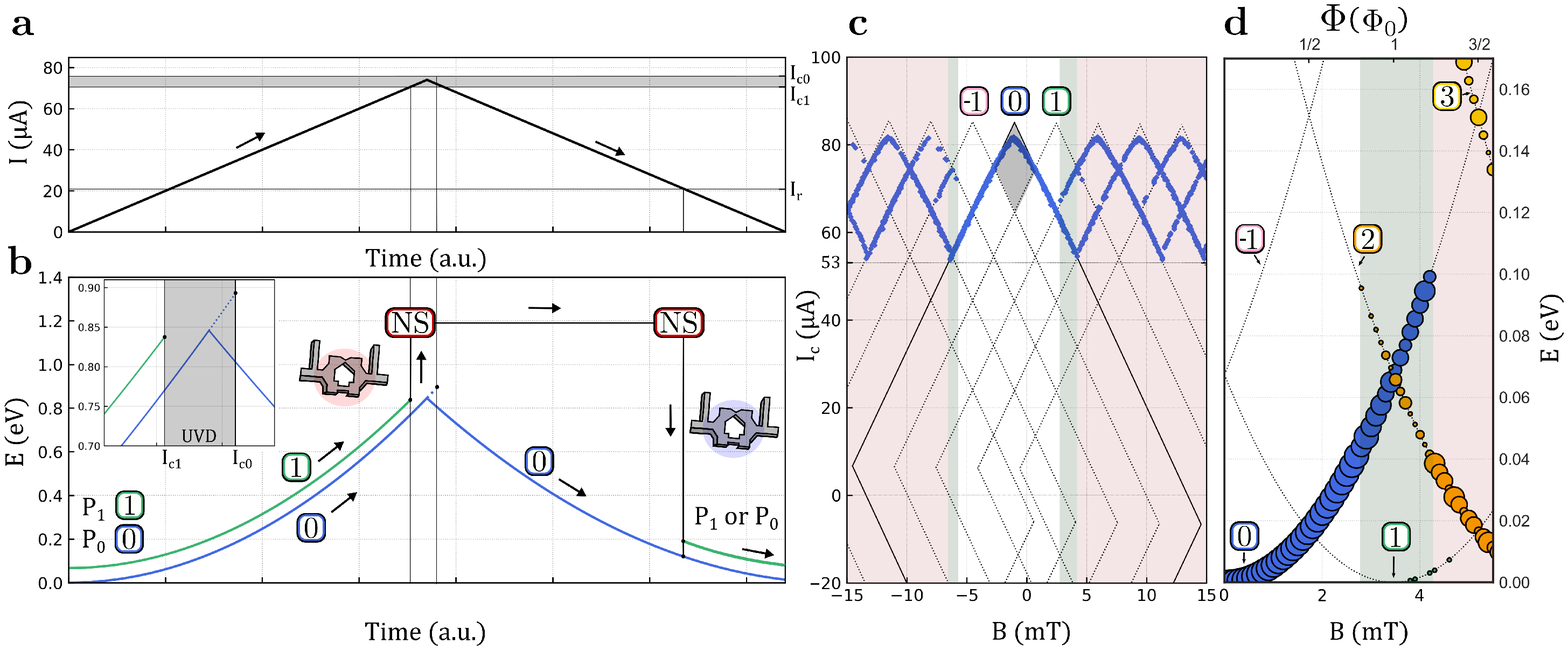}}
		\caption{(a) The time evolution of the applied current during one cycle of the preparation procedure under zero magnetic field. The current range between $I_{c0}$ and $I_{c1}$, indicated in gray, corresponds to the current range of the $n_v=0$ UVD. The retrapping current $I_r$ is field-independent and approximately $\sim$ 21.5 $\mu A$. (b) The energy stored in the NBSQUID for vorticity states 0 (blue) and 1 (green) during one cycle of the preperation procedure according to Equation \eqref{eq.energy}. The probability to measure these specific vorticity states is denoted by $P_0$ and $P_1$. At $I_{c1}$, vorticity state 1 ceases to exist and the SQUID switches to the normal state. As the applied current is decreased below the retrapping current the SQUID is frozen into vorticity state 0(1) with probability $P_{0(1)}$. The inset shows a zoom of the energy around the UVD current range. (c) The critical currents versus the applied magnetic field after first preparing the NBSQUID in vorticity state $n_v=0$ at $B=0$ mT. The dotted lines represent the vorticity diamonds generated by the model outlined in Section \ref{sec.model}. At each field value, we collected 7 critical currents. The $n_v=0$ UVD is indicated as a shaded gray fill. The shaded areas indicate the field values where the critical current is multivalued. (d) The energy of the different vorticity states of the NBSQUID as a function of the applied field (or equivalently, flux) at zero bias current. The size of the colored dots represents the probability to measure the critical current corresponding to the indicated vorticity state, extracted from the experimental data in panel c.}
		\label{fig2}
	\end{figure*}
	
	\begin{figure*}[ht!]
		\centering
		\subfigure{\includegraphics[width=\linewidth]{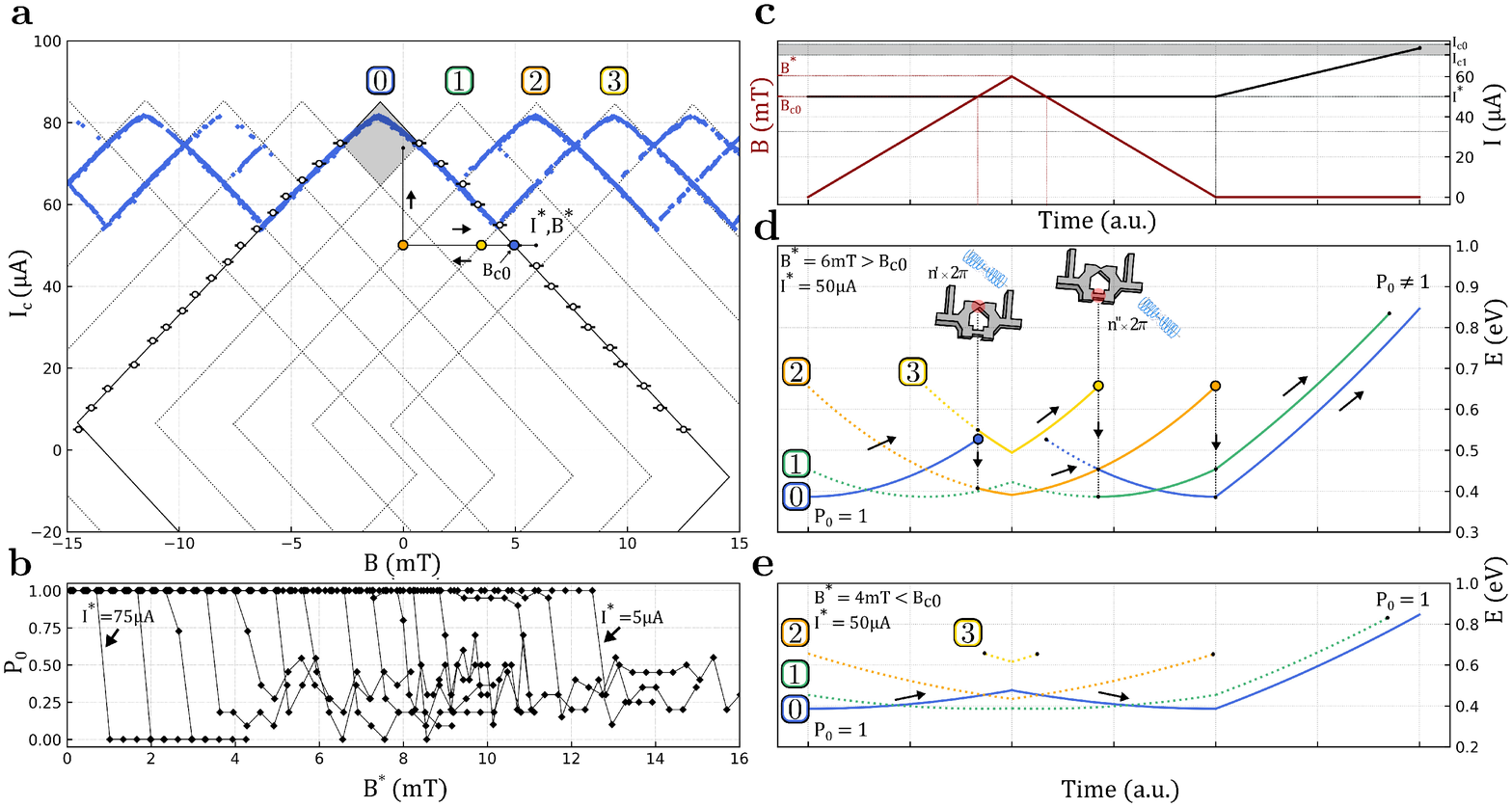}}
		\caption{(a) The critical currents versus the applied magnetic field after preparing the NBSQUID in vorticity state $n_v=0$ are indicated in blue. The open black circles with errorbar correspond to the experimentally obtained edge of the $n_v=0$ vorticity state using the measurement protocol as indicated by the black arrows and described in Section \ref{sec.lowbias}. (b) The probability to end up in vorticity state $n_v=0$ after sweeping to a particular $I^*$ and $B^*$. The curves correspond to bias currents $I^*$ ranging from $5$ $\mu A$ to $75$ $\mu A$. The field locations of the sudden jumps from $P_0=1$ to $P_0\neq1$ for a specific $I^*$ correspond to the right edge of the $n_v=0$ vorticity diamond in panel a. (c) The time evolution of the applied current (black) and field (red) during the measurement protocol as described in Section \ref{sec.lowbias}, for the case $B^*>B_{c0}$. The gray area denotes the current range of the $n_v = 0$ UVD at $B$ = 0 mT. (d) The energy stored in the NBSQUID as a function of the time for the vorticity states $n_v=0$, 1, 2 and 3 after initializing in $n_v=0$ and during the measurement protocol outlined in Section \ref{sec.lowbias}, for $I^*=50$ $\mu A$ and $B^*=6$ mT $> B_{c0}$. As $B^*=6$ mT exceeds $B_{c0}$ the NBSQUID's vorticity is altered by the means of a $n'\times 2\pi$ phase slip in the Dayem bridge SQUID arm. During readout, some of these vorticity states $n'$ cease to exist and the system switches to another vorticity state by a $n"\times 2\pi$ phase slip event in the nanobridge SQUID arm. (e) The energy stored in the NBSQUID as a function of the time for the vorticity states $n_v=0$, 1 and 2 after initializing in $n_v=0$ and during the measurement protocol outlined in Section \ref{sec.lowbias}, for $I^*=50$ $\mu A$ and $B^*<B_{c0}$. As $B^*$ never exceeds the critical field value $B_{c0}$ of vorticity state $n_v=0$, the system remains in vorticity state $n_v=0$.}
		\label{fig3}
	\end{figure*}

	\section{Critical current oscillations in a NBSQUID after initializing the vorticity state}\label{sec.initializing}
	
	Due to the stochastic nature of the freezing process, it is impossible to know the vorticity state of the system at the beginning of the $IV$-measurement. Therefore, no explicit information can be obtained about the phase dynamics occurring while sweeping the current. However, taking advantage of the statistics of the freezing process allows to prepare the NBSQUID into a particular vorticity state with a very high fidelity \cite{article_Murphy_A_2017}. In the preparation procedure, the concept of a `unique vorticity diamond' (UVD) is introduced. Inside the UVD there is only one stable vorticity state. In Figure \ref{fig1}a and Figure \ref{fig2}c, the UVD associated with vorticity $n_v=0$ is indicated as a dark gray shaded area. 
	\newline
	\newline
	To prepare the system in a specific vorticity state, one first needs to apply an external magnetic field corresponding to the UVD field range. For the $n_v=0$ diamond, $B$ = 0 mT is chosen. Then, a bias current of 74 $\mu A$, which leads to the normal state for all vorticity states, except for the one associated with the UVD is applied. Afterwards, the current is again reduced to 0 $\mu A$. This sequence is illustrated in Figure \ref{fig2}a: it shows the time evolution of the applied current during the preparation procedure under zero magnetic field. The current range indicated in gray corresponds to the current range of the $n_v=0$ UVD. The retrapping current $I_r$ is also indicated on the right axis. Its value is field-independent and approximately $\sim$ 21.5 $\mu A$. This process can also be looked at from an energy-standpoint, as shown in Figure \ref{fig2}b. The blue and green curves show the energy stored in the NBSQUID for the vorticity states as a function of time for $n_v=0$ and $n_v=1$ respectively. The time axes of panels a and b denote the same time, so that one can also indirectly interpret the curves in panel b as energy as a function of applied current. If at the starting time, the NBSQUID is in vorticity state $n_v=0$, applying a current above $I_{c1}$ but below $I_{c0}$ will not change this: the blue curve still exists. One can apply a current of 74 $\mu A$ and decrease it again without ever leaving the $n_v=0$ state. However, if the NBSQUID starts in the $n_v=1$ state, this is not true. For currents higher than $I_{c1}$, the $n_v =1$ energy state no longer exists and the SQUID consequently switches to the normal state (this can more clearly be seen in the inset). The SQUID then stays in the normal state, until the applied current is decreased again to the retrapping current $I_r$. At this point, the SQUID gets frozen into the superconducting state again, but due to the randomness of the freezing process it is not a priori known into which vorticity state. From the data shown in Figure \ref{fig1}a, the probability to end up in the $n_v=0$ state after a freezing process is extracted as $P_0\sim 0.99$ at $B$ = 0 mT. So even if the NBSQUID starts in another vorticity state than $n_v=0$, the probability to end up in the $n_v=0$ state remains high. To ensure a high fidelity of this preparation process,  the current is cycled $k = 5$ times between $I = 0$ $\mu A$ and $I = 74$ $\mu A$ at zero field ($B = 0$ mT). A zero voltage reading in the UVD of $n_v=0$ at $I=74$ $\mu A$ confirms that the state is indeed prepared correctly. (If this is not the case, $k$ is increased.) As such an experimental state preparation with 100 $\%$ fidelity is guaranteed. We observed that the prepared state remains stable at an applied bias current of 74 $\mu A$ for at least four hours.
	\newline
	\newline
	Figure \ref{fig2}c shows the measured critical current versus field $I_c (B)$ oscillations of the SQUID after first preparing in vorticity $n_v=0$ at $B=0$ mT. After initialization in the $n_v=0$ vorticity state, the bias current is set to zero, the applied magnetic field is changed to the value of interest and finally the critical current is obtained by performing an $IV$-measurement from $0$ $\mu$A to $90$ $\mu$A. The critical current at which the NBSQUID switches to the normal state is shown by the blue data points. This measurement procedure was repeated seven times at each field value. The dashed lines again result from the fit. Similar to Figure \ref{fig1}a and e the energy of the different vorticity states of the NBSQUID as a function of applied field, together with colored dots whose size represents their occurence are shown in Figure \ref{fig2}d.
	\newline
	\newline
	After the state preparation, the SQUID remains in vorticity state $n_v=0$ in the field range of $B_{n_v=0}\in[ B(I_{n_v=0}^{max} )-4.89 \text{ mT},B(I_{n_v=0}^{max} )+3.63 \text{ mT}]$, where $B(I_{n_v=0}^{max} )=-0.88$ mT is the magnetic field corresponding with the top vertex of the $n_v=0$ diamond. This can be seen from the single-valuedness of the measured critical current corresponding with the $n_v=0$ state in this field range, as shown by the white region in Figure \ref{fig2}c and d. For field ranges outside this interval the critical current is again multivalued, indicating that a hidden phase slip process altered the vorticity state during the measurement. There are two different distinct regimes observable, both for positive and negative field values. For field values inside the green region in Figure \ref{fig3}c and d vorticity $n_v=0$ is still observed, indicating the vorticity of the NBSQUID is sometimes altered by a hidden phase slip(s). While for the red indicated region vorticity $n_v=0$ is never observed, implying the vorticity is always changed by the means of a hidden phase slip(s).
	\newline
	\newline
	From the experimental data in Figure \ref{fig2}c and d it is apparent that for positive magnetic field values the $n_v=1$ vorticity state is barely observed. The notable abscence of the $n_v=1$ vorticity state can not be explained energetically, as vorticity the $n_v=1$ has the lowest energy state in the green region indicated in Figure \ref{fig2}d. This observation implies that upon leaving the $n_v=0$ vorticity state, it is more likely to have multiple phase slips occurring. It should be noted that this has been observed in MoGe nanowire SQUIDs \cite{article_belkin2015formation}. In this work, the authors concluded that there exists a regime in which paired phase slips are exponentially more likely to occur than a single phase slip. In this regime, the parity of the vorticity is thus conserved, which could be highly relevant for parity-protected qubits in future generation quantum computing applications \cite{article_douccot2012physical}. Finally, we have to remark that the observed vorticity states are not symmetric around the tip of the vorticity diamond. This indicates that there is a clear difference in phase slip event occurrence on both sides of the $n_v=0$ diamond. Indeed, when leaving the vorticity diamond through the upper right edge, this happens by means of a phase slip in the Dayem bridge arm as $\varphi_2=\varphi_{2c}$, see Figure \ref{fig1}d. When the diamond is exited through the upper left side, a phase slip occurs at the nanobridge SQUID arm as there $\varphi_1=\varphi_{1c}$ ( Figure \ref{fig1}c). Since the two arms' parameters (most importantly, their critical phase differences) differ, the observed asymmetry is not surprising. The subtle dependence of the phase dynamics on the geometry of the device and the exact value of the magnetic field when crossing the vorticity diamond can be used as a tool to study the metastability of a certain vorticity state at low bias currents not accessible through conventional $IV$-measurements.

	% the $n_v=1$ state is (almost) never observed when leaving the $n_v=0$ state trough the right diamond edge, while the $n_v=-1$ is observed upon leaving the $n_v=0$ state trough the left diamond edge (even though its occurrence is also less frequent than expected from energy considerations). 

	\section{The metastability of a vorticity state in a NBSQUID at low bias currents}\label{sec.lowbias}
	
	The question remains if the $n_v=0$ state remains metastable for field values outside the field-interval $B_{n_v=0}$. From a measurement of the critical current it is impossible to obtain this information, as this only reflects the vorticity value of the NBSQUID for bias currents exceeding $I=53$ $\mu A$. To gain more insight in the phase dynamics in the region of lower bias currents, we use a measurement protocol introduced in ref. \cite{article_Murphy_A_2017} to explore the metastability of the $n_v=0$ state and to reveal the occurrence of hidden phase slips. This procedure is also applicable to other vorticity states.
	\newline
	\newline
	In this measurement protocol (see the black arrows in Figure \ref{fig3}a), we first prepare the NBSQUID in the $n_v=0$ state at $B=0$ mT, using the procedure described in Section \ref{sec.initializing}. Subsequently, we fix the bias current to a value, $I^*$, and sweep the field towards a field value, $B^*$. Finally, we read-out the vorticity state by sweeping the field back to $B$ = 0 mT and the bias current to $I=74$ $\mu$A (a location within the $n_v=0$ UVD), where we can differentiate whether the NBSQUID is in the $n_v=0$ state or not by measuring the resistance: zero resistance corresponds to the $n_v=0$ state, while normal state resistance corresponds to another state state. This protocol performed after initializing the NBSQUID state is further illustrated in Figure \ref{fig3}c. In it, the time evolution of the applied magnetic field and the bias current are shown by red and black curves respectively. In this figure, $B^*$ is greater than $B_{c0}$, where $B_{c0}$ denotes the field at the right vorticity diamond edge at a current of $I^*$. The value of $I^*$ is 50 $\mu A$ in this figure and $B_{c0}$ is equal to 5 mT. $I_{c0}$ and $I_{c1}$  are the critical currents associated with vorticity states $n_v=0,1$ at the read-out field $B$ = 0 mT, such that the gray shaded area corresponds to the $n_v=0$ UVD. $I_r$ is the retrapping current.
	\newline
	\newline
	For different bias currents $I^*$ this measurement is performed 10 - 20 times. Figure \ref{fig3}b shows the probability to end up in the $n_v=0$ state after sweeping to a particular (in this case positive) field value $B^*$ at a fixed bias current $I^*$ and performing the read-out. Each trace contains a jump at a particular field value. The field value corresponding to the jump is indicated by an open circle for the different bias currents in Figure \ref{fig3}a and corresponds to the right edge of the vorticity diamond obtained from the fit. An analogous measurement for negative $B^*$ values results in values on the left edge of the $n_v=0$ vorticity diamond.
	\newline
	\newline
	For all bias currents the read-out indicates that the state remains 100 $\%$ in the $n_v=0$ state within the field range of the vorticity diamond, i.e. $B^*<B_{c0}$. This indicates that the NBSQUID remains in the $n_v=0$ state for the whole field and current phase space of the corresponding vorticity diamond (light gray shaded area in Figure \ref{fig1}a), while this is not necessarily the vorticity state lowest in energy. This is illustrated in Figure \ref{fig3}e, which shows the energy stored in the NBSQUID as a function of time for $I^*$ = 50 $\mu A$ for a $B^*$ value (4 mT) chosen inside of the $n_v=0$ vorticity diamond. The energies of the $n_v=0,1,2$ and $3$ states are shown using blue, green, orange and yellow curves respectively. As $B^*$ never exceeds the critical field value $B_{c0}$ of vorticity state $n_v=0$, the system remains in vorticity state $n_v=0$ even though other vorticity states have lower energies. This indicates a strong level of metastability of the vorticity state, implying that the phase slip rate is negligible.
	\newline
	\newline
	For bias currents exceeding $I=64$ $\mu A$, the probability to be in the $n_v=0$ state at read-out condition is zero after leaving the $n_v=0$ diamond. For these bias currents, the SQUID transits immediately to the normal state when leaving the $n_v=0$ diamond. Upon moving again to the read-out condition, the SQUID remains in the normal state as the retrapping current of $21.5$ $\mu A$ $<$ 53 $\mu A$. For bias currents below $I=53$ $\mu A$, the probability to be in the $n_v=0$ state at read-out condition is below 1 after leaving the $n_v=0$ diamond. This can be explained as follows. Consider leaving the $n_v=0$ diamond at a current below $I=53$ $\mu A$ by sweeping the field towards a value outside of the vorticity diamond edge, i.e. $B^*= 6 \text{mT} >B_{c0}$ (see Figure \ref{fig3}c). The energies for vorticity states $n_v=0, 1, 2, 3$ during this process are shown in Figure \ref{fig3}d by blue, green, orange and yellow curves respectively. It is clear that the SQUID exits the $n_v=0$ state once the edge of the vorticity diamond is crossed at a particular $(B,I)$-position, indicated by the blue dot labeled $B_{c0}$ in Figure \ref{fig3}a. As shown in our previous measurements at this particular $(B,I)$-position (Figure \ref{fig3}d, blue dot), the vorticity state of the NBSQUID is in this particular case changed to either $n_v=2$ or $n_v=3$  by a phase-slip process of $2\pi n'$ at the Dayem bridge side with a particular probability $P_{n_v'}$. Upon moving back to the read-out point, we leave the $n_v=2$ or $n_v=3$ state by a phase-slip process at the nanobridge side of $2\pi n''$, with a probability which can differ from $P_{n_v'}$. The positions where this occurs are marked by an orange ($n_v=2$) and yellow dot ($n_v=3$) in Figure \ref{fig3}a and d. Since the change in vorticity in going from the preparation point to the $(B^*,I^*)$-position is not necessarily the same as that in the `return trip', the vorticity at the read-out point can be changed from the prepared vorticity, which is reflected in the non-zero probabilities in the trace shown in Figure \ref{fig3}b.
	\newline
	\newline	
	\section{Validation with time-dependent Ginzburg-Landau simulations}
	\label{sec2tdgl}
	To validate the vorticity diamond model and gain insight into the phase dynamics, we performed time-dependent Ginzburg-Landau (tdGL) simulations of the NBSQUID. The behavior of the superconducting condensate is described by a complex-valued order parameter which is allowed to vary in time and space. We have used the tdGL equations for dirty superconductors \cite{kramer, gl_app}, where the order parameter is described by:

	\begin{align}
		\begin{split}
			& \frac{u}{\sqrt{1+\gamma^2\abs{\Psi}^2}}\left(\frac{\partial}{\partial t} + i\varphi + \frac{\gamma^2}{2}\frac{\partial \abs{\Psi}^2}{\partial t}\right)\Psi  \\
			& = (\grad - i \vb{A})^2\Psi + \left(1-\abs{\Psi}^2\right)\Psi.
		\end{split}
		\label{eq:gl1}
	\end{align}

\begin{figure*}[ht!]
		\centering
		\includegraphics[width=\textwidth]{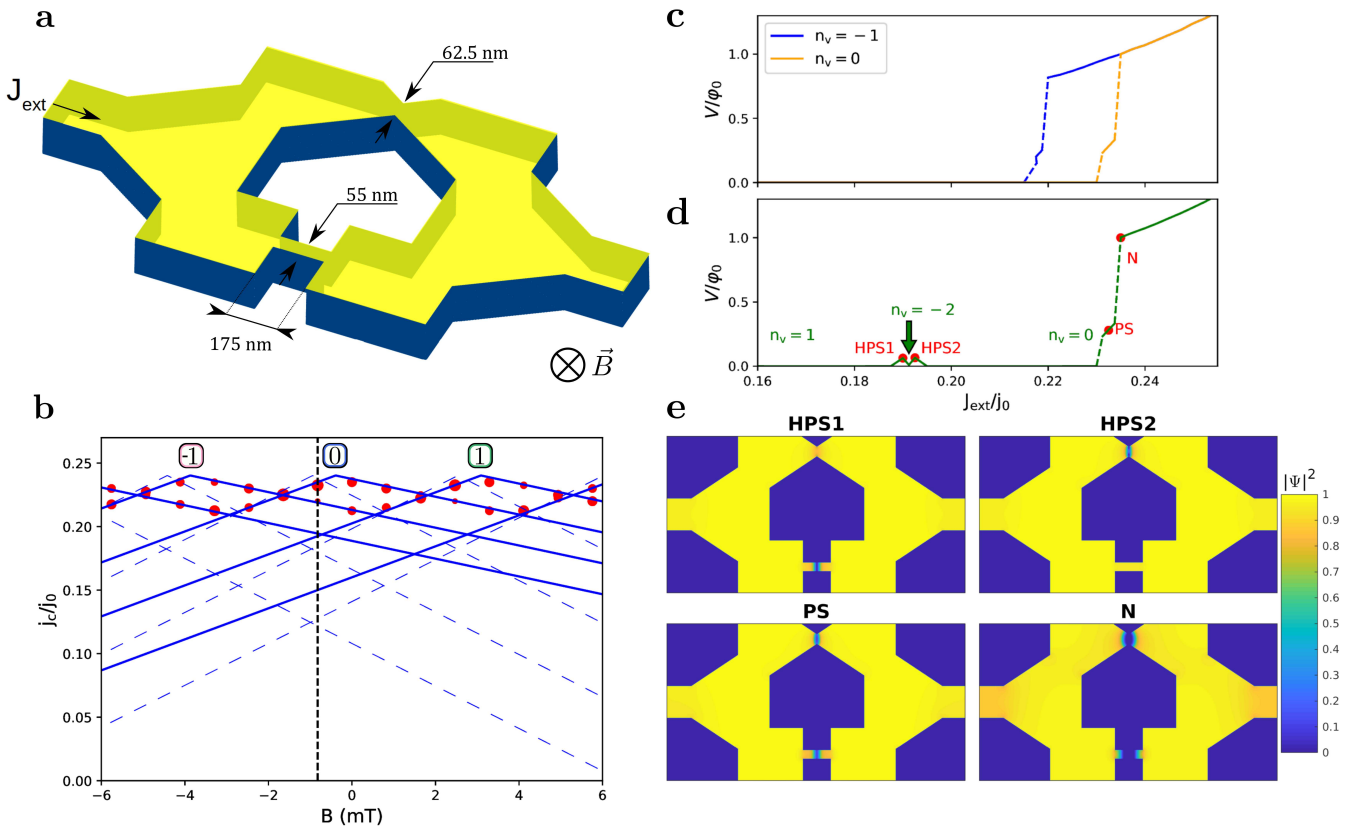}
		\caption{(a) The geometry of the simulated device and the orientation of the applied magnetic field (B) and current density ($J_{ext}$). (b) Simulated critical currents as a function of the applied magnetic field (red dots). The sizes of the dots are proportional to the probability of obtaining a particular value of the critical current. The solid blue lines represent vorticity diamonds from the simulation data. Increasing the width of the nanobridge by $\sim 10\%$ shifts the diamonds to a closer agreement with experimental data (dashed lines). (c,d) Selected current-voltage characteristics (\textit{IV}s) at the field marked by the dashed black line in panel b (-0.88 mT). Panel c shows two \textit{IV}s without hidden phase slips, starting from vorticity states $n_v=-1$ and $n_v=0$. While in panel d, starting from initial vorticity state $n_v=1$, two hidden phase slip events occur. The spikes related to hidden phase slips HPS1 and HPS2 lead to the vorticity states indicated in green. The transition to the normal state (N) is preceded by a running phase slip (PS). (e) Cooper-pair densities of the states marked with red dots in panel d. HPS1: three antivortices entering via HPS in the bottom bridge. HPS2: two antivortices leaving via HPS in the top bridge. PS: a continuously running phase slip, just before the transition to the normal state (N).}
		\label{fig4}
	\end{figure*}

	Here $u\approx 5.79$ is the ratio of the relaxation time for the amplitude and phase of the order parameter. $\vb{A}$ is the external magnetic vector potential and $\varphi$ is the electrostatic potential. $\gamma$ is a measure of the dirtiness of the sample, which characterizes the influence of the inelastic phonon-electron scattering on the condensate. Equation \eqref{eq:gl1} is solved self-consistently with the following equation for the scalar potential.
	\begin{equation}
		\laplacian{\varphi} = \div \Im\left[\Psi^* (\grad - i \vb{A})\Psi\right].
		\label{eq:gl2}
	\end{equation}
	An external current is applied as a boundary condition to the Poisson Equation \eqref{eq:gl2}. These equations are dimensionless, i.e.\ all lengths are measured in units of coherence length $\xi$, magnetic fields in units of the upper critical field $H_{c2}$, time in units of Ginzburg-Landau time $\tau_{GL} = \pi\hbar / 8k_BT_c u$, current densities in units of $j_0=\frac{3\sqrt{3}}{2}j_{dp}$, where $j_{dp}$ is the depairing current density, voltage in $\varphi_0 = \hbar / e^* \tau_{GL}$, etc.
	\\
	\\
	The equations are solved on a spatially discretized lattice, with 4 grid points per coherence length $\xi=10$~nm, implying that the simulation box of $2\text{ }\mu\text{m} \times 1$ $\mu$m is 800 by 400 pixels. The approach of link variables is used, where the order parameter $\Psi$ is defined on lattice nodes and the vector potential $\vb{A}$ on the links between them. The time step is chosen sufficiently small to guarantee the numerical stability of forward-time central-space integration scheme for Equation \eqref{eq:gl1}. For the numerical simulations, an idealized geometry was used. The Dayem bridge is 62.5 nm wide in the narrowest point, the nanobridge is 55 nm $\times$ 175 nm, as shown in Figure \ref{fig4}a.

	\vspace{10pt}
	For each value of applied magnetic field 7 current sweeps were executed. First, the order parameter was initialized randomly to simulate freezing in from the normal state, resulting in spontaneous nucleation of vortex-antivortex pairs. (Anti)Vortices either leave the loop through the external boundary or enter the hole, contributing to the net vorticity. The stochasticity of the nucleation process leads to different initial vorticity states for the current sweep. A video visualizing the evolution of the Cooper-pair density for this nucleation process can be found in the supplementary material. Subsequently, the system was evolved deterministically as the applied current was ramped up. The multivalued critical current as a function of the applied field follows the vorticity diamond model as indicated by the red dots in Figure \ref{fig4}b. The size of these dots correspond to the probability of obtaining the particular critical current value. Direct examination of the phase of the order parameter just before the transition to the resistive state shows that the vorticity agrees with the diamond number. The Little-Parks oscillation period $\Delta B$ matches exactly the experimental value of 3.48 mT. The top vertices of diamonds from simulations are less shifted from $B=0$ mT compared to the experiment, indicating a smaller difference in the critical phases of the bridges. Like in the experiment, switching currents show rounding at the top vertices of the vorticity diamonds, reflecting a nonlinear C$\Phi$R in this current range. Larger observed differences in the slopes of left and right branches indicate that bridges in the simulation exhibit different inductances. This is likely due to the more pronounced bridge asymmetry in the idealized geometry versus the case of the real sample. As the critical phase angle of a bridge is proportional to its length and the inductance to its aspect ratio \cite{article_dausy2021impact}, increasing the width of the nanobridge by $\sim 10\%$ shifts the diamonds to a closer agreement with experimental data. This adjustment is indicated by the blue dashed line in Figure \ref{fig4}b.
	\\
	\\
	The performed numerical simulations can offer a direct insight into the phase dynamics by following the current-voltage (\textit{IV}) characteristics, for example the \textit{IV} curve obtained at the field value indicated by the black dashed line in Figure \ref{fig4}b. There are two possible cases in these \textit{IV} characteristics,  hidden phase slip(s) can occur or no hidden phase slip(s) occur. Figure \ref{fig4}c shows the case where no hidden phase slip(s) take place, the system starts in vorticity $n_v =$ -1 (0) and by increasing the current the system will transit to a dissipative state indicated in blue (orange). Figure \ref{fig4}d shows the case with two hidden phase slips events. In this figure, the SQUID was frozen into vorticity $n_v$ = 1. By increasing the current the edge of the vorticity diamond ($J_{ext}$ = 0.1900) of the $n_v$ = 1 vorticity is reached. At this point a hidden phase slip indicated by a small spike (HPS1) occurs, this phase slip brings the device to vorticity $n_v$ = -2. By further increasing the current value the edge of this state is reached ($J_{ext}$ = 0.1925) and a second hidden phase slip (HPS2) occurs, resulting in vorticity $n_v$ = 0. Further increasing the current will lead to a switching to the normal state, as was the case in Figure \ref{fig4}c. The evolution of the Cooper-pair density of the events labeled in Figure \ref{fig4}d, are visualized in Figure \ref{fig4}e. As HPS1 (2) occurs on the left (right) side of the vorticity diamond a discontinuity in the nano (Dayem) bridge is observed, while the transition to the normal state (N) is preceded by a running phase slip (PS) in both bridges.

	\section{Conclusion}
	We investigated the metastability and phase slip dynamics of the different energy states associated with the vorticity or winding number of a MoGe nanobridge SQUID. By utilizing the unique vorticity diamond the system can be initialized in a specific vorticity state. Based on the measurement conditions (the freezing process), we demonstrated that the system is already rendered in a metastable energy state. This could prove of interest for future technological advancements, e.g. time resolved pulses for memory applications, or preparation in higher vorticity states for signal enhancement similar to ref. \cite{opremcak2018measurement}. By controlling the initial state and determining the final state by measuring the $I_c(B)$ oscillations this metastability was examined. Moreover, we are not limited to the region where the SQUID transits to the normal state. At low bias currents the used measurement protocol uncovered the hidden phase slip regime. These phase slip(s) could be associated with and explained by their corresponding energy landscape. Not only can we determine in which specific arm of the SQUID this phase slip happens, for certain cases one can exactly determine which phase slip occurs (e.g. Figure \ref{fig3}a at 5 mT). Due to the translational periodicity of the $I_c(B)$ oscillations the analysis for different initial vorticity states will be analogous. Complementary tdGL simulations validated the vorticity diamond model and demonstrated that the prepared NBSQUID remains trapped within the whole vorticity diamond region. Further simulations showed that phase slips occured at certain field values with at least multiplicity two. The small discrepancy between the experiment and simulations can be explained due to the idealized SQUID geometry.

	\vspace{2ex}
	\noindent
	\begin{acknowledgements}
		This work is supported by the Research Foundation–Flanders (FWO, Belgium), Grant No. G0B5315N, as well as by the COST action NanoCoHybri (CA16218).
	\end{acknowledgements}
	\vspace{2ex}

	\newpage
	\bibliographystyle{unsrt}
	\bibliography{ref.bib}

\end{document}